\documentclass[prd,aps,twocolumn,groupedaddress,showpacs]{revtex4}
\begin{document}
\input{psfig.sty}

\title{On the Circular Orbit Approximation for \\ 
Binary Compact Objects 
In General Relativity}

\author{Mark Miller}
\affiliation{238-332 Jet Propulsion Laboratory, 4800 Oak Grove Drive,
Pasadena, CA  91109}

\date{\today}

\begin{abstract}                                                        %
 
One often-used approximation in the study of binary compact objects
(i.e., black holes and neutron stars)
in general relativity is the instantaneously circular orbit 
assumption.  This approximation has been used extensively, from 
the calculation of innermost circular orbits to the 
construction of initial data for numerical relativity
calculations.
While this assumption
is inconsistent with generic general relativistic astrophysical
inspiral phenomena
where the dissipative effects of gravitational radiation 
cause the separation of the compact objects to decrease in time, it
is usually argued that the timescale of this dissipation is much longer
than the orbital timescale so that the approximation of circular
orbits is valid.  Here, we quantitatively analyze this approximation using
a post-Newtonian approach that includes terms up to order 
$({Gm/(rc^2)})^{9/2}$
for non-spinning particles.  By calculating the evolution of equal mass
black hole / black hole binary systems
starting with circular orbit configurations and comparing them to 
the more astrophysically relevant
quasicircular solutions, we show that a minimum initial separation
corresponding to at least $6$ ($3.5$) orbits 
before plunge is required in order to
bound the detection event loss rate in gravitational wave detectors
to \hbox{$< \: 5\%$~($20\%$).}  In addition, we show that the detection
event loss rate is $ > \: 95\%$ for a range of initial separations that
include all modern calculations of the innermost circular orbit (ICO).

\end{abstract}


\maketitle

\section{Introduction}                                                  %
\label{sec:introduction}                                                %

The construction of astrophysically relevant initial data is a crucial
prerequisite to using numerical relativity as a predictive
tool in gravitational wave astronomy, i.e. in
producing gravitational waveform templates to be used in searches for
signals in 
modern interferometric gravitational wave detectors such as LIGO,
VIRGO, GEO600, and TAMA300.
After all, the comparison between results from a numerical relativity
evolution code and any observed astrophysical phenomena would be meaningless
unless the initial data used in the 
numerical evolution corresponds to an astrophysically realistic situation.
In particular, numerical relativists are interested in
constructing astrophysically realistic initial data
corresponding to the late stages of the inspiral process of compact objects
(black holes and neutron stars), so that the 
transition from the final few orbits to the subsequent coalescence
and formation of the final merged object can be numerically simulated using
the full nonlinear theory of general relativity to accurately detail these
highly nonlinear phenomena.  To date, all initial data constructed
for this purpose employ some sort of circular orbit assumption in the
construction of the initial data.  For instance, many 
quasiequilibrium studies assume the existence of a helical
Killing vector field in the construction of configurations that
correspond to compact object binaries that are instantaneously in
circular motion~\cite{Baumgarte98b,Baumgarte98c,Bonazzola97,Gourgoulhon01,
Marronetti:1999ya,Shibata98,Teukolsky98,Yo01,Duez00,Duez02,
Gourgoulhon:2001ec,Grandclement:2001ed,Friedman02}.  
Other methods employ turning point
techniques to arrive at circular orbit 
configurations~\cite{Cook94,Baumgarte00a}.
It is typically argued that because the 
timescale of the gravitational radiation is longer than the orbital
timescale of the binary, the assumption of quasiequilibrium is a good
one.  However, as the orbital separation of the compact objects 
decreases during the evolutionary progression of the binary, the
timescale of the gravitational radiation process increases while
the orbital timescale decreases.  In other words, quasiequilibrium
assumptions become increasingly inaccurate 
as the compact objects get closer.  It is
therefore imperative to assess, in an astrophysically meaningful way,
at what point during the inspiral process the quasiequilibrium 
approximation breaks down.  In this paper, we present a method 
for calculating a lower bound on the error of the quasiequilibrium 
approximation for inspiraling binaries.  This method is 
demonstrated by calculating this lower bound for equal mass 
black hole / black hole binaries and neutron star / neutron star
binaries.

One specific aspect of the quasiequilibrium approximation is the
circular orbit assumption, and it is precisely this aspect which we
analyze in this paper.  
Our analysis is carried out within a 
post-Newtonian non-spinning point particle approximation.
Using initial
data corresponding to two particles in circular orbit, we solve the
post-Newtonian equations of motion to find the resulting
evolution of the binary system.  The purpose of this calculation
is twofold.  First, we seek to quantitatively assess the circular orbit
assumption in the construction of initial data in full general 
relativity.  We do this by computing the correlation function of
the gravitational wave signal produced from the evolution starting
with particles in circular orbit with the gravitational wave
signal from the true
quasicircular evolution (i.e., the unique, ``fully circularized'' 
evolution of the binary system within the post-Newtonian 
approximation).  This gives us a meaningful measure of the error
introduced by the circular orbit assumption in the initial data.
Second, we wish to have a baseline with which to compare fully 
general relativistic calculations (i.e., calculations in numerical
relativity) using initially circular orbiting
compact objects as initial data.

We find that, as expected, the initial circular orbit assumption
becomes progressively worse as the initial separation of the compact 
objects decreases.  Specifically, we find that for equal
mass black hole
binaries, the initial separation must exceed a separation such 
that $6.3$ ($3.5$) orbits remain before plunge
in order to bound the detection event loss rate in gravitational
wave detectors to $< \: 5\%$ ($20\%$).
This method is less efficient in calculating the lower bound of
the error induced by the initial circular orbit approximation
for neutron stars, due to the fact that hydrodynamical effects
will become important earlier in the evolution of the binary.  As a result,
the lower bound in the detection event loss rate for equal mass neutron
stars that we calculate are never larger than several percent.

The remainder of the paper is organized as follows.   In 
section~\ref{sec:quasicircular}, we describe the 
post-Newtonian equations of motion for two particles.  We solve
these equations for very large initial separation, thus producing
the unique, ``fully circularized'' solution corresponding to the 
last 1000 quasicircular orbits of the binary inspiral. 
In section~\ref{sec:circular}, we formulate initial data corresponding to 
compact objects in initially circular orbits for various initial 
separations and compare the resulting gravitational wave signals to the 
true quasicircular gravitational wave signals.  
In section~\ref{sec:nr}, we introduce a local definition of eccentricity
which we find useful for comparisons to numerical relativity calculations,
and compute the expected eccentricity
as a function of initial separation that one can expect to find in
numerical relativity calculations using initial data corresponding to
initially circular orbiting compact binaries.
We compare these results with a fully consistent numerical relativity
calculation of a neutron star binary inspiral using circular orbit
initial data.  In the conclusions, section~\ref{sec:conclusions}, we
comment on the implications our results have on the 
astrophysical relevance of current innermost circular
orbit (ICO) calculations.

\section{Quasicircular Binary Evolution}
\label{sec:quasicircular}

The general relativistic equations of motion
for non-spinning point particles can be written in a post-Newtonian
expansion as
\begin{eqnarray}
\frac {d^2 \vec{x}} {dt^2} & = & - \frac {m}{r^2} \hat{n}  +
   \frac {m}{r^2} \left [ \hat{n} (A_{1PN} + A_{2PN} + 
                                   A_{3PN} + \cdots) \right. + \nonumber \\
   & & \left.  \dot{r} \vec{v} (B_{1PN} + B_{2PN} + B_{3PN} + \cdots
   ) \right ] + \nonumber \\
   & & \frac {8}{5} \eta \frac {m}{r^2} \frac {m}{r} \left [ \dot{r} 
      \hat{n} (A_{2.5PN} + A_{3.5PN} + A_{4.5PN} + \cdots ) - 
      \right. \nonumber \\
   & & \left. \vec{v} (B_{2.5PN} + B_{3.5PN} + B_{4.5PN} + \cdots )
      \right ],
\label{eq:pneom}
\end{eqnarray}
where $\vec{x} = \vec{x}_2 - \vec{x}_1$ is the relative separation of
the particles, $\vec{v} = \vec{v}_2 - \vec{v}_1$ is the relative 
velocity between the particles, $r = |\vec{x}|$, $\hat{n} = \vec{x}/r$, 
$m = m_1 + m_2$, $\eta = m_1 m_2 / m^2$, and $\dot{r} = dr/dt$.
The post-Newtonian expansion is carried out in powers
of $\epsilon \sim m/r \sim v^2$ (here, we have set $G=c=1$).

Of the non-radiative terms in the expansion (i.e., terms that are
of integer order powers in the post-Newtonian expansion), only 
the 1PN and 2PN terms have been completely 
determined~\cite{Blanchet98,Itoh01,Pati02};
the 3PN terms have been calculated up to one 
numerical parameter~\cite{Damour2000_Poinc,Blanchet01a}. 
Of the radiative terms (e.g., n.5PN terms)
in the expansion, only the 2.5PN~\cite{Blanchet98,Itoh01,Pati02}
and 3.5PN~\cite{Pati02} terms have been completely 
determined.  Employing an energy and angular momentum
balance technique, the 4.5PN terms have been determined modulo
$12$ free ``gauge'' parameters~\cite{Gopakumar97}.  
Here, we numerically
solve Eq.~\ref{eq:pneom} using all post-Newtonian terms up
to and including the first radiation reaction terms (i.e. up to and
including the 2.5PN terms) for orbiting binaries.  In order to obtain
information regarding the error introduced by truncating the 
post-Newtonian expansion, we also calculate solutions which
include the radiative 3.5PN and 4.5PN terms.  We hereafter
refer to this system of equations (i.e. up to and including the
2.5PN terms, plus the 3.5PN and 4.5PN terms) as the 
``4.5PN'' equations, acknowledging that we are not
including the conservative 3.0PN and 4.0PN terms.  We have found
empirically that the solutions to these 4.5PN equations of motion
are highly insensitive to the 12 ``gauge'' parameters in the 4.5PN
terms~\cite{Gopakumar97}.  More specifically, the differences in the 
solutions to the 4.5PN equations of motion when the 12 parameters
are randomly varied between $-100$ and $100$ are orders of magnitude 
smaller than the differences in solutions obtained where the
4.5PN terms are dropped altogether.  
We take the binary to be in the $x$-$y$ plane, which we 
coordinatize by the usual
polar coordinates $(r,\phi)$.  Once the initial values of
$r$, $\phi$, $\dot{r}$, and $\dot{\phi} \equiv d\phi/dt$ are
set, the equations of motion completely specifies the solution.

\begin{figure}
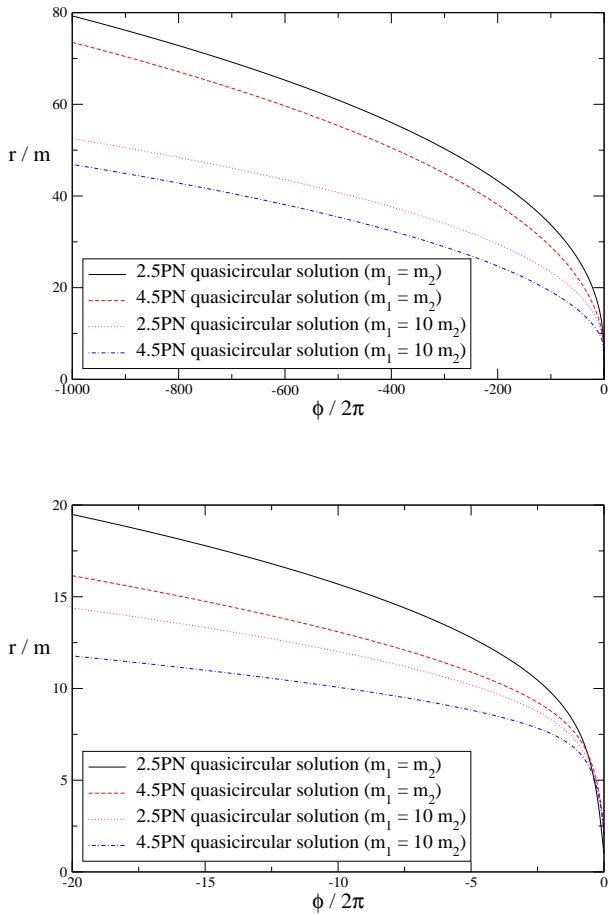


\psfig{figure=r_vs_phi.eps,width=8cm}

\vspace{1.0cm}

\psfig{figure=r_vs_phi_2.eps,width=8cm}

\caption{The orbital separation $r$ as a function of the number
of orbits for the quasicircular solution of the 2.5PN and 4.5PN
equations of motion.
We show the equal mass ($m_1 = m_2$) case
along with the case of mass ratio $m_1/m_2 = 10$.  The top panel displays
the last 1000 orbits, the bottom panel displays the last 20 orbits.
}
\vspace{0.0cm}
\label{fig:r_vs_phi}
\end{figure}

As shown in~\cite{Lincoln90}, the effect of the radiation reaction terms
in the post-Newtonian equations of motion for highly 
separated compact binaries $(r > 1000 \: m)$ is to circularize
the orbit of the binary.  We refer to this state, and its subsequent
solution to the post-Newtonian equations of motion as the
unique (up to rotations) 
``quasicircular'' solution.
For an arbitrary binary separation $r_i$, 
initial data which corresponds to
this quasicircular orbit scenario is given at 2PN order
(see~\cite{Lincoln90})
as $\dot{r}_i = 0$ with the square of the initial angular velocity
$\dot{\phi}_i$ satisfying the cubic equation
\begin{eqnarray}
0 & = & 
{(r_i^2 \dot{\phi}_i^2)}^3 -
\frac {m}{r_i} {(r_i^2 \dot{\phi}_i^2)}^2 +
(3 - \eta) {(\frac {m}{r_i})}^3 {(r_i^2 \dot{\phi}_i^2)} - \nonumber \\
 & & (15 + \frac {17}{4} \eta + 2 \eta^2){(\frac {m}{r_i})}^5.
\end{eqnarray}
Since this initial data does not give
a true quasicircular evolution when considering equations
of motion of order 2.5PN or higher, we use it as initial data for
numerical integrations starting with large initial 
separations $r_i$ (typically $r_i > 100 \: m$), 
and allow the equations of motion to fully circularize
the orbit, after which we refer to the solution as {\it the} 
unique (up to spatial rotations) quasicircular solution.

\begin{figure}
\vspace{0.0cm}
\hspace{0.0cm}
\psfig{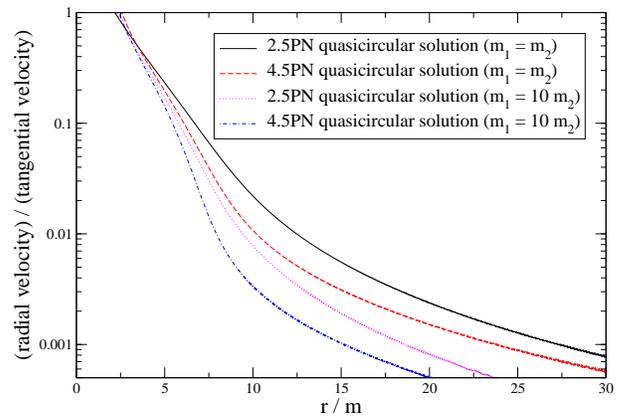}
\caption{The ratio of the radial velocity $|\dot{r}|$ to the
tangential velocity $r \dot{\phi}$ is plotted as a function
of the binary separation $r$ for the quasicircular
solution for both the 2.5PN equation of motion and the
4.5PN equation of motion.  We show the equal mass ($m_1 = m_2$) case
along with the mass ratio $m_1/m_2 = 10$ case.
}
\vspace{0.0cm}
\label{fig:qe_speed_rat}
\end{figure}

In Fig.~\ref{fig:r_vs_phi}, we plot the binary separation
$r$ as a function of $\phi$ for the quasicircular solution to the
2.5PN and 4.5PN equations of motion 
for both the equal mass ($m_1 = m_2$) case,
as well as the case where the mass ratio $m_1/m_2$ is $10$.
We easily use enough resolution so that the truncation error, if 
plotted as error bars on all curves in this paper, would 
be smaller than the thickness of the curves.
We see that the 4.5PN 
solution displays a somewhat 
weaker radiation damping than the 
2.5PN solution, in the sense that if one starts from
a specific initial separation, there are more orbits until plunge
for the 4.5PN solution as compared to the 2.5PN solution.
For example, the bottom panel of Fig.~\ref{fig:r_vs_phi}
shows that for a circularized binary with orbital separation 
$r = 15 \: m$, the 2.5PN solution evolves for approximately 9 orbits
until plunge (which, for definiteness, we define here as $r = 2 \: m$), 
whereas the
4.5PN quasicircular solution evolves for approximately 16 orbits
until plunge.
In Fig.~\ref{fig:qe_speed_rat}, the ratio of the radial velocity
$|\dot{r}|$ to the tangential velocity $r \dot{\phi}$ as a function
of separation $r$ for the 2.5PN and 4.5PN quasicircular solutions is plotted, 
again for both the equal
mass case and the mass ratio $m_1/m_2 = 10$ case.  In both cases,
we see that for separations
$r > 4 \: m$, the ratio of the radial velocity to the 
tangential velocity is smaller for the 4.5PN solution than for the 
2.5PN solution; once the separation reaches $4 \: m$, 
the merger is a fraction of an orbit away for both the 2.5PN and 4.5PN 
solutions.

\section{Circular Initial Data and Subsequent Binary Evolution}
\label{sec:circular}

The particular aspect of the quasiequilibrium approximation
used in constructing initial data for 
binary systems of compact objects in numerical relativity 
that we analyze here is the
circular orbit assumption.  In practice, this assumption can take the
form of an assumption of the existence of a
helical Killing vector field; alternatively, 
turning point methods can be used to find
approximate circular orbit configurations.  
The assumption of a circular orbit configuration as 
initial data in general relativity,
while fully consistent from a mathematical point of view, is 
certainly questionable from an astrophysical point of view.  
A casual inspection of Figs.~\ref{fig:r_vs_phi}~and~\ref{fig:qe_speed_rat}
reveals that a
circular orbit assumption in the initial data for a realistic
binary system (which we effectively define as a 
fully circularized binary; we ignore
here any capture scenarios and/or dynamically driven scenarios 
where the binary could have high eccentricities all the way
down to the plunge phase) will be increasingly inaccurate 
as the initial separation
decreases.
We note that considerations of computational resources 
alone will limit accurate
simulations performed by numerical relativity codes in the coming decades 
to timescales of roughly $10$ orbital periods.
From Fig.~\ref{fig:r_vs_phi}, this translates to initial binary separations
$r_i < 20 \: m$.
The relevant question is therefore:
at what initial orbital separation $r_i$ does the circular orbit
approximation break down?
In other words, will gravitational waveforms
produced from an evolution code using
initial data corresponding to compact objects in exactly circular orbit be,
in some sense, close enough to waveforms produced from
the more astrophysically relevant quasicircular case? 
Here, we answer this question quantitatively using the PN approximation
described in the previous section.

\begin{figure}
\vspace{0.0cm}
\hspace{0.0cm}
\psfig{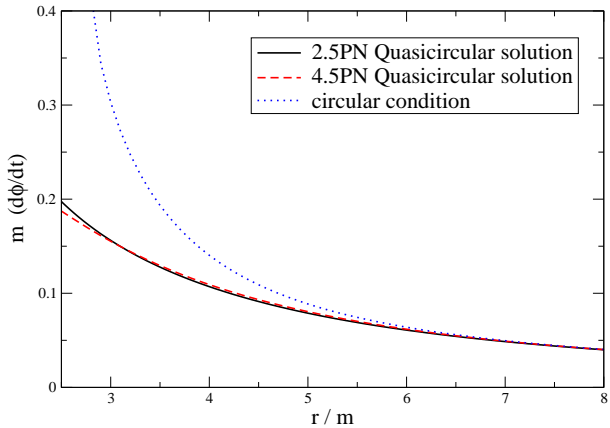}
\caption{The angular velocity $\dot{\phi} = d{\phi}/dt$ is plotted as
a function of binary separation $r$ for the quasicircular solution
to the 2.5PN and 4.5PN equations of motion.  Also shown is the 
angular velocity corresponding to instantaneously 
circular orbits ($\dot{r} = 0$ and $\ddot{r} = 0$, see Eq.~\ref{eq:rddot})
at separation $r$.
}
\vspace{0.0cm}
\label{fig:circ_phidot}
\end{figure}

For an arbitrary initial separation $r_i$, we construct 
initial data at time $t_i = 0$ 
corresponding to exact circular orbits by requiring the vanishing of
the first and second time derivatives of the separation, 
$\dot{r}_i \equiv (dr/dt)|_{t=0}$ and
$\ddot{r}_i \equiv (d^2r/dt^2)|_{t=0}$.  
Assuming $\dot{r}_i = 0$, the expression for the instantaneous 
radial acceleration $\ddot{r}_i$ is given by the equations of
motion, Eq.~\ref{eq:pneom}, as
\begin{equation}
\ddot{r}_i = r_i {\dot{\phi}_i}^2 + \frac {m}{r_i^2} 
   (-1 + A_{1PN} + A_{2PN} + A_{3PN} + \cdots).
\label{eq:pnrdotdot}
\end{equation}
We note that the radiation reaction
terms (the n.5PN terms) do not enter in the prescription for
circular data in the post-Newtonian approximation.  Using the 
completely determined post-Newtonian $A_{1PN}$ and $A_{2PN}$ 
terms (see, e.g.,~\cite{Pati02}),
the expression for the instantaneous radial acceleration,
Eq.~\ref{eq:pnrdotdot}, can be written explicitly as
\begin{eqnarray}
\ddot{r}_i & = r_i {\dot{\phi}_i}^2 + \frac {m}{r_i^2} 
   \left [  \right. & -1 - (1 + 3 \eta) r_i^2 {\dot{\phi}_i}^2 +
   2 (2 + \eta) \frac {m}{r_i} -  \nonumber \\
 & & \eta (3 - 4 \eta) r_i^4 {\dot{\phi}_i}^4 + 
      \frac {1}{2} \eta (13 - 4 \eta) m r_i {\dot{\phi}_i}^2 - \nonumber \\
 & &  \frac {3}{4} (12 + 29 \eta) {(\frac {m}{r_i})}^2.
\label{eq:rddot}
\end{eqnarray}
We see that, to 2PN order, the equation $\ddot{r}_i = 0$ (assuming
$\dot{r}_i$ is also zero) becomes a
quadratic equation for the square of the initial angular 
velocity $\dot{\phi}_i$.  The initial circular orbit
assumption,
$\dot{r}_i = 0$ and $\ddot{r}_i = 0$,
thus completely specifies the initial configuration for any arbitrary
initial separation $r_i$.  The initial angular velocity
$\dot{\phi}_i$ for this circular condition
is plotted in Fig.~\ref{fig:circ_phidot} as a function of 
initial separation $r_i$ for the equal mass binary case.  
For comparison, the angular velocity
for the 2.5PN and 4.5PN quasicircular solution is also shown.
We see that the circular orbit approximation induces an 
artificial increase in the angular velocity as compared
to the more astrophysically relevant quasicircular solutions.  
Differences between the two become apparent at 
roughly $r = 6 \: m$, and by $r = 3 \: m$ the angular velocity
for the circular orbit approximation is double that of the 
quasicircular solutions.

\begin{figure}
\vspace{0.0cm}
\hspace{0.0cm}
\psfig{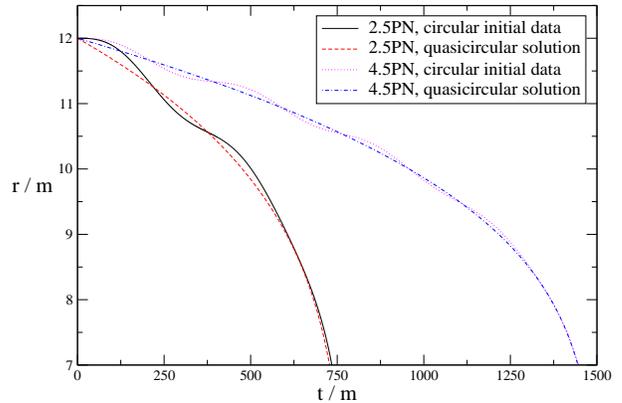}
\caption{The evolved binary separation $r$ as a function of time $t$ is 
plotted for evolutions starting with circular initial data, with the
initial separation $r_i = 12 \: m$ for the equal mass case ($m_1 = m_2$).
Both 2.5PN solutions and 4.5PN solutions are plotted.  For comparison,
the quasicircular solution is also shown for both the 2.5PN and 4.5PN case.
}
\vspace{0.0cm}
\label{fig:r0_qc_12m}
\end{figure}

In Fig.~\ref{fig:r0_qc_12m}, we plot solutions to the 2.5PN and 4.5PN
equations of motion, using the initially circular conditions
$\dot{r}_i = 0$ and $\ddot{r}_i=0$ as initial data, with the 
initial separation $r_i = 12 \: m$.  Here, we show the
equal mass case, $m_1 = m_2$.  For comparison, the quasicircular
solutions to the 2.5PN and 4.5PN equations of motion, where we
have set $t=0$ when $r = 12 \: m$, are also shown.  We see that the
circular orbit approximation in the initial data induces oscillations
(with a period of roughly $380 \: m$)
in the separation $r$, but that the evolutionary track of 
the two evolutions are similar.
We see that for both the 
2.5PN and 4.5PN circular initial data cases,
the orbit is slightly eccentric, with $t=0$ corresponding to
an apastron point in the orbit.  Similar surveys of solutions
starting with different initial binary separation parameter $r_i$ reveal
the expected result that the oscillations in the evolved separation $r$
using circular initial data
decrease monotonically as the initial separation $r_i$ increases;  
i.e., the ``error'' induced by assuming circular initial data
decreases with increasing initial binary separation $r_i$.  The fact
that the circular orbit approximation induces an eccentricity in the
orbit is not unexpected.  It was shown in~\cite{Mora02} that allowing
for eccentric orbits provided a better match between post-Newtonian
turning point calculation results and numerical relativity turning
point calculation results.

\begin{figure}
\vspace{0.0cm}
\hspace{0.0cm}
\psfig{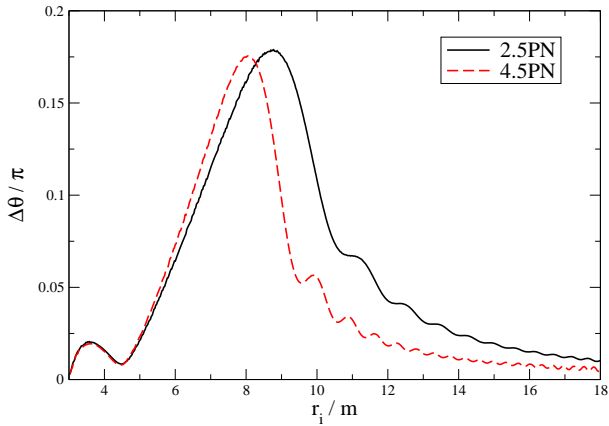}
\caption{The time averaged phase difference, 
$\Delta \theta$ (Eq.~\ref{eq:dtheta}), between the 
quasicircular solution and the solution using instantaneously circular
initial data for evolutions starting with initial 
binary separation $r_i$.  The solid curve corresponds to solutions
to the 2.5PN equations of motion and the dashed curve corresponds to
solutions to the 4.5PN equations of motion.
}
\vspace{0.0cm}
\label{fig:r0qc_phase}
\end{figure}

In order
to get an idea of the phase errors induced by the circular orbit
approximation, we calculate $\Delta \theta$, defined by
\begin{equation}
\Delta \theta = \sqrt{
   \frac {\int^{t_f}_0 dt \: {( \theta_{c}(t) - \theta_{qc}(t) )}^2}{t_f} },
\label{eq:dtheta}
\end{equation}
which is a measure of the time averaged phase difference between the 
two solutions to the post-Newtonian equations of motion;  $\theta_{c}(t)$
corresponding to the phase angle of the solution using circular orbit
initial data and $\theta_{qc}(t)$ corresponding to the phase angle of the 
quasicircular solution, where each solution has an initial 
separation $r_i$ ($\Delta \theta$ is therefore an implicit function 
of initial separation $r_i$).  Here, we introduce a cutoff time
$t_f$, which we choose to be the time when either of the solutions
reaches a binary separation of $r = 3 \: m$, since finite size
effects become important for separations smaller than $r = 3 \: m$
for black holes (recall that in harmonic coordinates, the horizon
of a static non-rotating black hole is located at $r = M$, where M is
the mass of the black hole).  In Fig.~\ref{fig:r0qc_phase}, we plot
$\Delta \theta$ as a function of initial binary separation $r_i$ in
the equal mass black hole case for both the 2.5PN and 4.5PN solutions.
We see that a maximum is reached for initial binary separations of
$r_i = 8.8 \: m$ and $r_i = 8.1 \: m$ for the 2.5PN and 4.5PN solutions,
respectively.  These separations correspond to roughly $1.5$ orbital 
periods before the final plunge.

\begin{figure}
\vspace{0.0cm}
\hspace{0.0cm}
\psfig{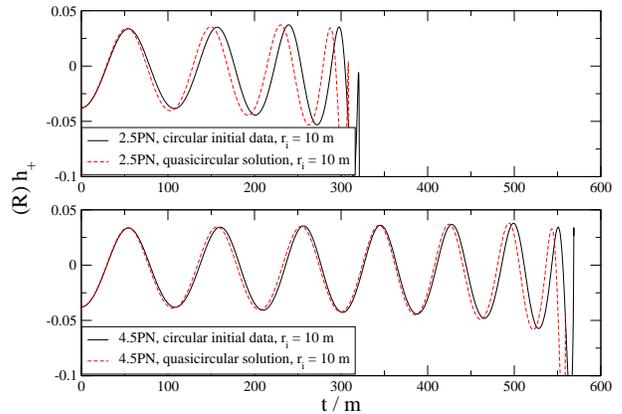}
\caption{The ``plus'' (+) polarization gravitational waveform for
coalescing, equal mass binaries.  The top panel shows the radiation
from solutions of the 2.5PN equations of motion and the bottom
panel shows the radiation from solutions of the 4.5PN equations of motion.
Both panels show the waveform for the solution starting with 
circular initial data and the quasicircular solution.  The initial 
separation in each case is $r_i = 10 \: m$.  The observation direction
is in the orbital plane, with $\Theta = \pi/2$ and $\Phi = 0$.
}
\vspace{0.0cm}
\label{fig:grwave_10m}
\end{figure}

From a gravitational wave detection point of view, the relevant
measure of how ``close'' the solutions starting with circular
initial data are to the more astrophysically realistic
quasicircular solutions is determined by examining the gravitational
wave signal associated with each solution. We use the
post-Newtonian formalism presented 
in~\cite{Epstein75,Wagoner76,Turner78,Lincoln90}, which give
the two polarization states $h_+$ and $h_\times$ for our solutions as
a function of observer distance $R$ and observation
directions $\Theta$ and $\Phi$.  In Fig.~\ref{fig:grwave_10m}, we show
gravitational waveforms for solutions to the post-Newtonian 
equations of motion using initial data starting at $r_i = 10 \: m$.  We 
see explicitly that the 2.5PN solution plunges within approximately
two orbits, while the 4.5PN solution plunges 
within approximately four orbits (one orbital period is
twice the period of the waveform).  We also see that the phase of the waveform
corresponding to the circular initial data slightly lags the phase
of the more astrophysically realistic quasicircular solution in both
the 2.5PN and 4.5PN cases.  The effect is less pronounced 
in the 4.5PN case.  

To quantify the difference between the gravitational waveform 
$h_c(t)$ obtained from circular
initial data and the gravitational waveform $h_{qc}(t)$ 
obtained from the quasicircular solution, we define the
correlation $C[h_c(t),h_{qc}(t)]$ of the waveforms as
\begin{eqnarray}
\lefteqn{C[h_c(t),h_{qc}(t)]  = } \nonumber \\
 &  & \max_\tau \left \{
   \frac { \int^{t_f}_{0} h_c(t) h_{qc}(t - \tau) \: dt}
      {\sqrt{ (\int^{t_f}_0 {(h_c(t))}^2 \: dt)
             (\int^{t_f}_0 {(h_{qc}(t - \tau))}^2 \: dt) } } \right \}.
\label{eq:corrdef}
\end{eqnarray}
Notice that the quantity inside the braces in the definition of
the correlation $C[h_c(t),h_{qc}(t)]$ depends on both the lag time $\tau$
and the cutoff time $t_f$ for the time integrations.  In our case, we
specify the cutoff time $t_f$ as the earliest time
when either of the waveforms
corresponds to 
some ``final'' binary separation $r_f$.  For black hole binaries,
we expect tidal effects to become important as $r < 3 \: m$, so we
set $r_f = 3 \: m$.  For neutron star equal mass binaries, hydrodynamical
effects will become important (depending on details of the equation
of state) at around $r = 9 \: m$, in which case we set $r_f = 9 \: m$.  
Once
the cutoff time $t_f$ is determined, the quantity inside the
braces in Eq.~\ref{eq:corrdef} is a function of the lag time $\tau$;  
the final correlation $C[h_c(t),h_{qc}(t)]$ is determined by varying 
the lag time $\tau$ and finding the maximum value of the expression
in braces.
It is important to note that by introducing a cutoff of the
gravitational waveform (which is necessary in a practical sense,
due to the fact that the post-Newtonian approximation we are using
breaks down when tidal effects become important), we are computing
a {\it lower bound} of the error induced by using circular initial
data for inspiral simulations;  a more realistic calculation of the
error would include all of the
nonlinear phenomena occurring {\it after} tidal effects become important,
which we are necessarily neglecting.

\begin{figure}
\vspace{0.0cm}
\hspace{0.0cm}
\psfig{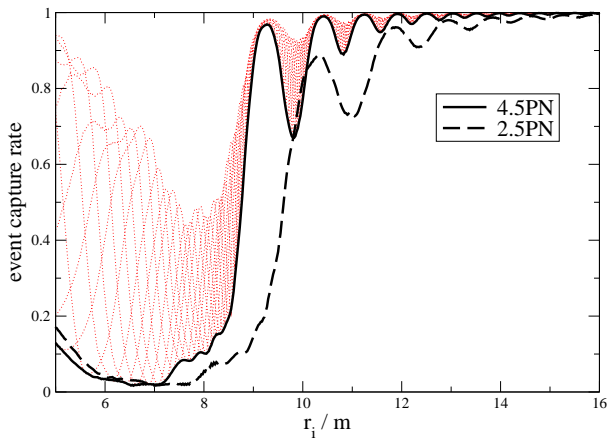}
\caption{The event capture rate, $C[h_c(t),h_{qc}(t)]^3$
(where $C[h_c(t),h_{qc}(t)]$ is defined
in Eq.~\ref{eq:corrdef}),
of 4.5PN gravitational waveforms using circular
initial data as a function of initial binary separation $r_i$.
The gravitational waveform observer is
located in the orbital plane $\Theta = \pi/2$, with
observer polar angles $\Phi = 0$, $\pi/8$, $2\pi/8$, $3\pi/8$,
$4\pi/8$, $5\pi/8$, $6\pi/8$, and $7\pi/8$
shown as thin dotted curves.  The thick
solid curve labeled ``4.5PN''
is the minimum correlation for all observation polar angles $\Phi$.
This same minimum is plotted for the 2.5PN case and is
labeled ``2.5PN''.
}
\vspace{0.0cm}
\label{fig:2.5corr}
\end{figure}

In Fig.~\ref{fig:2.5corr}, we plot the cube of the correlation function 
$C[h_c(t),h_{qc}(t)]$, Eq.~\ref{eq:corrdef}, between 
the waveform computed using
circular 
initial data and the waveform of the quasicircular solution as a function
of initial binary separation $r_i$, where we have set
$r_f = 3 \: m$.  This gives us an estimate of
the lower bound of the event loss rate induced by using 
circular initial data instead of the more astrophysically 
correct quasicircular initial data in the case where the compact 
objects are black holes.  For each initial separation
$r_i$, we can find the minimum of the correlation over all observation
polar angles $\Phi$, and take this to be the measure of the lower
bound of the error induced by the circular orbit approximation 
in the initial data.  This curve is 
labeled ``4.5PN'' in Fig.~\ref{fig:2.5corr}.
The oscillations in this curve, which have a period of slightly more
than one additional orbit in the evolution of the binary, correspond
precisely to additional oscillations in the $r$ vs. $t$ curves
(see Fig.~\ref{fig:r0_qc_12m}).  That is, as the initial
binary separation $r_i$ increases, more oscillations in the
$r$ vs. $t$ curve are allowed before the binary plunges, and
the oscillations in Fig.~\ref{fig:2.5corr} occur as a result.  
These oscillations are also evident in the time averaged
phase difference, $\Delta \theta$ (Fig.~\ref{fig:r0qc_phase}), and
occur for the same reason.

\begin{figure}
\vspace{0.0cm}
\hspace{0.0cm}
\psfig{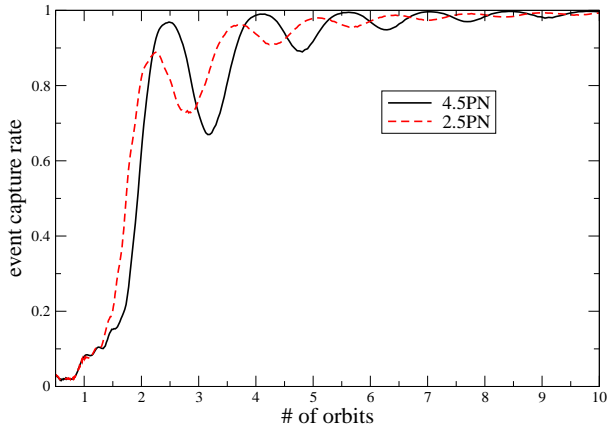}
\caption{
The event capture rate, $C[h_c(t),h_{qc}(t)]^3$
(where $C[h_c(t),h_{qc}(t)]$ is defined
in Eq.~\ref{eq:corrdef}),
of gravitational waveforms from inspiraling black holes 
($r_f = 3 \: m$) using circular
initial data as a function of the number of
orbits until plunge (i.e., until $r = 2 \: m$).  
The observer is
located in the orbital plane $\Theta = \pi/2$, and the
event capture rate is minimized over all polar angles $\Phi$
(see, e.g., Fig.~\ref{fig:2.5corr}).  Results using both the 4.5PN 
and 2.5PN equations of motion are shown.
}
\vspace{0.0cm}
\label{fig:corrorbit}
\end{figure}

We see from Fig.~\ref{fig:2.5corr} that, for any
given initial separation $r_i$, the correlation 
between the waveform obtained from
circular initial data and the waveform obtained from the quasicircular
solution is larger for the 4.5PN solutions than for the 2.5PN 
solutions.  However, recall from Fig.~\ref{fig:r_vs_phi} that,
for a given initial separation $r_i$, there
is quite a large difference in the number of orbits before final plunge
between the 2.5PN case and the 4.5PN case.  From the standpoint of
numerical relativity, the number of orbits before plunge is perhaps
the more important criterion in parameterizing the orbital 
separation of the binary.  In Fig.~\ref{fig:corrorbit}, we recast the
results in Fig.~\ref{fig:2.5corr}, plotting the cube of the
correlation function
$C[h_c(t),h_{qc}(t)]$, Eq.~\ref{eq:corrdef}, between
the waveform computed using
circular
initial data and the waveform of the quasicircular solution as a function
of the number of orbits until final plunge.
We see that the profile of this event capture rate is similar for 
both the 2.5PN and 4.5PN cases, the major difference between the two
being the phase of the oscillations.  This gives us confidence that
the post-Newtonian approximation has converged sufficiently to provide
us with an accurate estimate of the lower bound of the error induced
by the circular orbit assumption in the construction of
initial data.  The results of Figs.~\ref{fig:2.5corr}~and~\ref{fig:corrorbit}
are summarized in tabular form in Table~\ref{tab:eventloss}.  We can see
that using circular initial data with configurations that have $2$ orbits
until plunge or less will result in a $50\%$ or more event loss rate, and 
that one must start with configurations of $r_i > 10 \: m$ (with
more than $3.5$ orbits until plunge) in order to have a chance 
of bounding the event loss rate to below $20\%$.

\begin{table}
\begin{tabular}{|c|c|c|}  \hline \hline
\hspace{0.0cm} { Event loss rate} \hspace{0.0cm}  &
\hspace{0.0cm} {minimum initial } \hspace{0.0cm}  &
\hspace{0.0cm} {minimum initial } \\ 
\hspace{0.0cm} { } \hspace{0.0cm}  &
\hspace{0.0cm} {separation $r_i$ (2.5PN) } \hspace{0.0cm}  &
\hspace{0.0cm} {separation $r_i$ (4.5PN) } \\ \hline \hline
{1\%} &
   {$15.7 \: m$} {[9.8 orbits]} &
   {$13.8 \: m$} {[11.9 orbits]} \\ \hline
{5\%} &
   {$12.7 \: m$} {[4.7 orbits]} &
   {$11.6 \: m$} {[6.3 orbits]} \\ \hline
{20\%} &
   {$11.3 \: m$} {[3.1 orbits]} &
   {$10.0 \: m$} {[3.5 orbits]} \\ \hline
{50\%} &
   {$9.6 \: m$} {[1.7 orbits]} &
   {$8.8 \: m$} {[1.9 orbits]} \\ \hline
{80\%} &
   {$9.2 \: m$} {[1.5 orbits]} &
   {$8.6 \: m$} {[1.7 orbits]} \\ \hline \hline
\end{tabular}
\vspace{0.0mm}
\caption{The minimum initial binary separation $r_i$ for specific event
loss rates due to
errors in the initial data introduced by the circular orbit approximation
for black holes ($r_f = 3\:m$), as predicted by the 2.5PN and
4.5PN equations of motion.  In brackets are 
the number of orbits remaining until plunge, starting at this
initial binary separation (see 
Figs.~\ref{fig:2.5corr}~and~\ref{fig:corrorbit}).
}
\label{tab:eventloss}
\end{table}

For neutron stars, we expect tidal effects to become important
at binary separations $r$ of between $7\:m$ and $9\:m$, depending
on the equation of state of the nuclear matter.  As a result, the
point particle approximation breaks down much earlier for 
neutron star models than for black hole models.  This 
will necessarily result in a corresponding decrease in
sensitivity when determining the errors induced by using
circular orbit initial data.  However, this does not 
necessarily imply that the errors induced by
assuming initially circular orbits is less for neutron star
binaries than those for black hole binaries at similar separations, but 
instead is simply an artifact that this particular model
(spinless point particle model) is worse for neutron star
binaries than for black hole binaries; after all, we are only
computing lower bounds to these errors.

\begin{figure}
\vspace{0.0cm}
\hspace{0.0cm}
\psfig{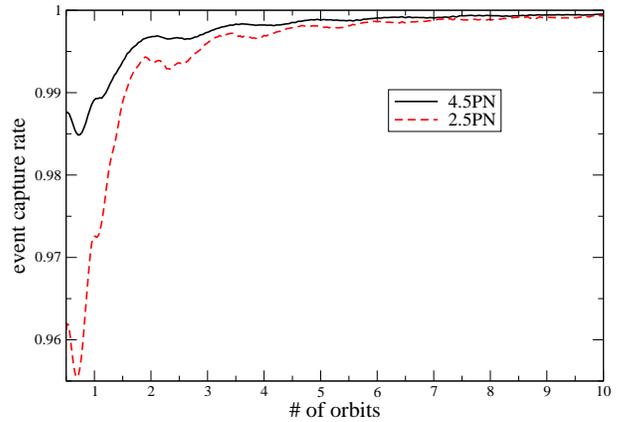}
\caption{
The event capture rate, $C[h_c(t),h_{qc}(t)]^3$
(where $C[h_c(t),h_{qc}(t)]$ is defined
in Eq.~\ref{eq:corrdef}),
of gravitational waveforms from inspiraling neutron stars 
($r_f = 9 \: m$) using circular
initial data as a function of the number of
orbits until separation $r = r_f = 9 \: m$, at which point
hydrodynamical effects will become important.
The observer is
located in the orbital plane $\Theta = \pi/2$, and the
event capture rate is minimized over all polar angles $\Phi$
(see, e.g., Fig.~\ref{fig:2.5corr}).  Results using both the 4.5PN
and 2.5PN equations of motion are shown.
}
\vspace{0.0cm}
\label{fig:NScorr}
\end{figure}

In Fig.~\ref{fig:NScorr}, we set the
cutoff separation $r_f$ to $9\:m$ (i.e., assuming neutron
star binaries) and plot the cube of the
correlation function $C$, Eq.~\ref{eq:corrdef}, between the
waveform computed using circular
initial data and the waveform of the quasicircular solution as a function
of initial binary separation $r_i$.  As expected, we 
observe a notable decrease in sensitivity to errors induced
by assuming initially circular data as compared to the 
black hole case ($r_f = 3\:m$, 
see Figs.~\ref{fig:2.5corr}~and~\ref{fig:corrorbit}).  In the case of
equal mass binary neutron stars, the lower bound on the event loss rate
that we compute is lower than $2\%$.

\section{Circular Initial Data: Comparing Post-Newtonian and Full Numerical
Relativity Calculations}
\label{sec:nr}

Numerical relativity has progressed to the point where stable, multiple
orbit numerical simulations of compact objects 
are now possible.  In~\cite{Miller03b}, we present fully general
relativistic simulations of binary neutron stars using conformally
flat, quasiequilibrium initial data corresponding to an equal mass,
corotating binary.  The construction of the initial data, in addition
to being a solution of the constraint equations 
of general relativity,  assumed the
existence of a timelike helical Killing vector field.  That is, the
binary is assumed to be instantaneously in a circular orbit.  We wish
to compare the numerical relativity calculations 
obtained in~\cite{Miller03b} with the
post-Newtonian formulation from the previous sections.
One difficulty in performing any such comparison is due to the fact
that each calculation was performed in different
coordinate systems.  The post-Newtonian calculations are performed
in a harmonic gauge ($g^{\alpha \beta} {\Gamma^{\mu}}_{\alpha \beta} = 0$),
whereas numerical relativity simulations typically use a variety of
other (both local and elliptic-type) gauge conditions to specify the
coordinates.  In particular, the relativistic calculations of binary
neutron stars presented in~\cite{Miller03b} used a quasi-isotropic
coordinate system; the initial spatial slice is
conformally flat, but the subsequent evolution of the full Einstein
field equations drives the 3-metric away from conformal flatness.
Nevertheless, we can attempt a comparison between the harmonic coordinate
post-Newtonian calculations with the quasi-isotropic
coordinate numerical relativity calculation by using a simple
coordinate transformation that takes the single, stationary, non-rotating
black hole solution of the Einstein equations written in harmonic 
coordinates~\cite{Cook97b} to the same solution 
written in isotropic coordinates.  Let $(t_{har},r_{har})$ and 
$(t_{iso},r_{iso})$ be the (time,radial) coordinate pair for
the harmonic and isotropic coordinatizations, respectively.
They are related by
\begin{eqnarray}
r_{har} & = & r_{iso} + \frac {m^2}{4 r_{iso}} \nonumber \\*
t_{har} & = & t_{iso} + 4 m \ln{ \left( \frac {2 r_{iso} - m}
   {2 r_{iso} + m} \right) }.
\label{eq:hariso}
\end{eqnarray}
We use this relationship to map the harmonic coordinate separation of the 
post-Newtonian calculations done in harmonic coordinates to the
isotropic separation of the fully general relativistic calculations 
in~\cite{Miller03b}
done in (quasi-) isotropic coordinates.

\begin{figure}
\vspace{0.0cm}
\hspace{0.0cm}
\psfig{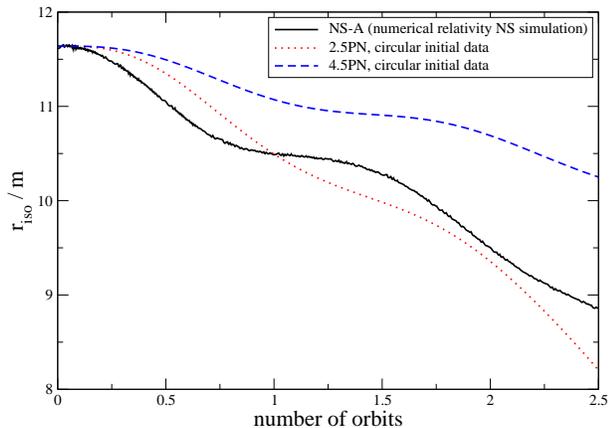}
\caption{The isotropic separation of binary evolution calculations
as a function of the number of orbits.  Each calculation uses
a circular orbit assumption as initial data.  The solid curve
refers to a fully consistent numerical relativity calculation of
binary neutron stars using conformally flat, quasiequilibrium initial
data (NS-A from reference~\cite{Miller03b}).
The dotted and dashed curves correspond to 2.5PN and 4.5PN
calculations, respectively, starting with the same initial
isotropic separation as that of NS-A, assuming initially circular
orbits.
}
\vspace{0.0cm}
\label{fig:gr_compare}
\end{figure}

In Fig.~\ref{fig:gr_compare}, we plot the isotropic, center of mass separation
from the numerical relativity calculation NS-A of 
reference~\cite{Miller03b} as a function of the number of
orbits.  For comparison, we calculate the same quantity using
the 2.5PN and 4.5PN equations of motion from the previous sections, starting
with the same isotropic binary separation as that of NS-A and using 
the initially circular orbit assumption.  The frequency
of the oscillations in the separation is slightly different
for the numerical relativity calculation NS-A as compared to the
post-Newtonian calculations.  There are many possible explanations for
this.  First of all, the calculations are done in 
different coordinate systems.  While an attempt has been made
to match the coordinates used to measure the 
binary separation (i.e., Eq.~\ref{eq:hariso}) in the two cases,
the differences in the angular and/or time coordinates could 
explain the difference.  Second, the numerical relativity 
calculation NS-A contains finite size effects, whereas the
post-Newtonian approximation does not.  Thirdly, the neutron
stars in the 
initial data used for the NS-A numerical relativity calculation
contain an unphysically high amount of spin, since the co-rotation
assumption was used in its construction.  The post-Newtonian
approximation we are using does not take spin into account.  Thus,
any spin-orbit coupling in the NS-A calculation will not be present in
the post-Newtonian approximation.  Finally, the numerical errors
in the calculation NS-A still could be large enough to explain
the difference.

\begin{figure}
\vspace{0.0cm}
\hspace{0.0cm}
\psfig{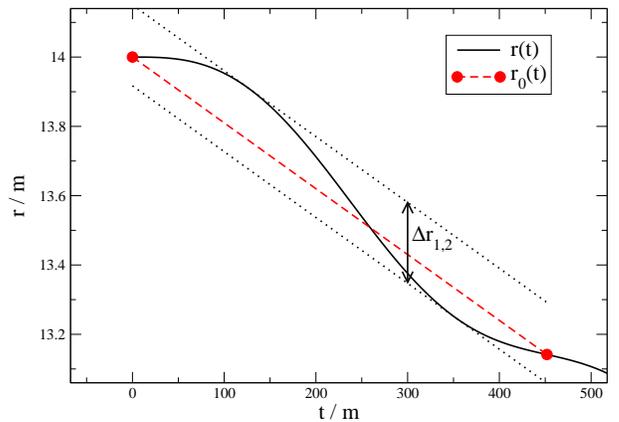}
\caption{An example of the definition of ${\Delta r}_{1,2}$,
used in the definition of the generalized
eccentricity, Eq.~\ref{eq:eccentric}.  Here,
$r(t)$ represents the 2.5PN solution using circular initial data
at an initial binary separation of $r_i = 14 \: m$.  The general
apastron points are denoted by circles, and ${r_0}(t)$ is the line
connecting the generalized apastron points.
}
\vspace{0.0cm}
\label{fig:eccentric}
\end{figure}

However, the {\it magnitudes} of the oscillations 
in the separations
shown in Fig.~\ref{fig:gr_compare}, which are a measure
of the eccentricity of the orbits, are comparable.  
Notice that the separation $r$ is a monotonically decreasing
function of time for all three calculations presented in 
Fig.~\ref{fig:gr_compare}.  We note that this is universally
true for post-Newtonian calculations using circular initial
data.  We therefore generalize the concept of an apastron point
(typically defined as a local maximum of separation $r$) for
monotonically decreasing binary separation functions r(t)
as follows:  if $r(t)$ is monotonically decreasing, then we define
a {\it generalized} apastron point to be those points where
$dr/dt$ attain a local maximum.  We note that the numerical
relativity calculation NS-A, as well as the 2.5PN and 4.5PN 
solutions using circular initial data, are initially ($t=0$ 
corresponds to $0$ orbits in Fig.~\ref{fig:gr_compare}) at 
generalized apastron points.  Given any two consecutive generalized
apastron points of a separation function 
$r(t)$ at, say,  times $t_1$ and $t_2$, we can 
define the eccentricity $e_{1,2}$ as follows.  Let 
${\langle r \rangle}_{1,2}$ be the time averaged value
of the binary separation $r$ between the two 
generalized apastron points:
\begin{equation}
{\langle r \rangle}_{1,2} = \frac {\int^{t_2}_{t_1} dt \: r(t)}
   {(t_2 - t_1)}.
\end{equation}
Let ${r_0}(t)$ be a linear function of time that intersects 
the points $r(t_1)$ and $r(t_2)$.
We define ${\Delta r}_{1,2}$ to be the range of $r(t)-{r_0}(t)$ 
within the interval $t = [t_1,t_2]$ (see Fig.~\ref{fig:eccentric}).
The generalized eccentricity $e$ is then defined as
\begin{equation}
e = \frac { {\Delta r}_{1,2} } {2 \: {\langle r \rangle}_{1,2}}.
\label{eq:eccentric}
\end{equation}
The advantages of this particular definition of the generalized 
eccentricity are twofold.  In the first place, Eq.~\ref{eq:eccentric}
reduces to 
the usual definition of eccentricity in the 
Newtonian limit.  Secondly, it is defined in terms of 
quantities local to the binary, as opposed to quantities
defined at spatial infinity.  This makes it particularly 
useful in numerical relativity, where invariant measures
of the binary separation can be straightforwardly computed.

\begin{figure}
\vspace{0.0cm}
\hspace{0.0cm}
\psfig{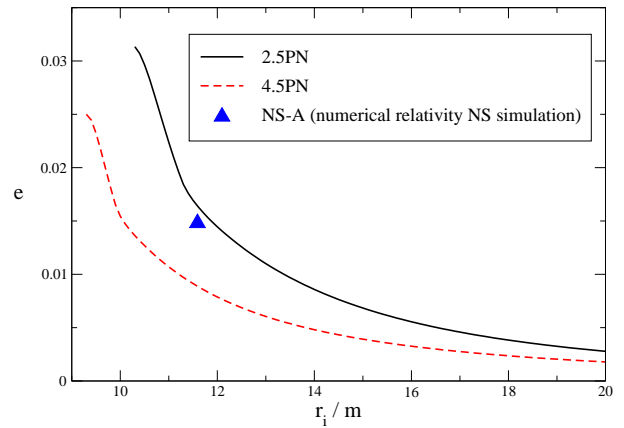}
\caption{The generalized eccentricity $e$, Eq.~\ref{eq:eccentric},
as a function of initial separation $r_i$
for initially circular orbiting equal mass
binaries.   Shown are results for both the 2.5PN and 4.5PN
equations of motion.  Also shown as the solid triangle is the
eccentricity $e$ of the numerical relativity simulation of
binary neutron stars
NS-A from reference~\cite{Miller03b}.
}
\vspace{0.0cm}
\label{fig:evsr}
\end{figure}

In Fig.~\ref{fig:evsr}, we plot the generalized eccentricity $e$
(Eq.~\ref{eq:eccentric}) for initially circular orbiting equal mass
binaries as a function of initial binary separation $r_i$.  
The first generalized apastron point used in each calculation
is at the initial time $t=0$.  Shown are calculations using
both the 2.5PN and 4.5PN equations of motion.
At small values of 
initial separation $r_i$ (less than $r_i = 10.07 \: m$ for the 2.5PN equations
of motion and $r_i = 9.18 \: m$ for the 4.5PN equations of 
motion) the solutions r(t) have monotonically decreasing 
first derivatives $dr/dt$, and thus do not exhibit multiple 
generalized apastron points (one could define the generalized
apastron points in this case to be points where $d^2r/dt^2$ obtain local 
maxima, thereby further generalizing the definition of
apastron points, but we have not done so here).  For comparison, 
we show as a solid triangle in Fig.~\ref{fig:evsr} the results
from numerical relativity simulation NS-A from reference~\cite{Miller03b}
of equal mass
binary neutron stars using conformally flat, quasi-equilibrium
initial data.  In order to make the
comparison more meaningful, we use Eq.~\ref{eq:hariso} to 
relate the quasi-isotropic coordinate separation from 
the numerical relativity calculation to the harmonic coordinates
used by the post-Newtonian calculation (although the difference between
the two are quite small for $r_i = 11.6 \: m$).  As can be seen, the
eccentricity observed in the numerical relativity calculation 
NS-A lies between the 2.5PN and 4.5PN eccentricity predictions.  Again,
this comparison must be viewed as preliminary, as there still
may exist significant numerical errors (i.e., boundary and/or
truncation errors) in the numerical 
relativity simulations.

\section{Conclusions}
\label{sec:conclusions}

In this paper, we have performed an analysis of the circular
orbit approximation which is often used in the analysis of binary compact
objects in general relativity.  Using astrophysically relevant measures
of the errors induced by the circular orbit approximation,
we have shown that for equal mass
binary black holes, the approximation has completely broken down 
for binary separations of $r \leq 8.5 \: m$, which correspond to separations
where roughly $2$ orbits or less remain before the final plunge.  
We conclude that numerical relativity evolution codes 
using circular orbit initial data configurations will be 
required to perform at least $3.5$ orbits (or more, depending on
the accuracy required, see Table~\ref{tab:eventloss}) in order to
produce astrophysically meaningful waveforms to be used as 
search templates in modern interferometric gravitational wave detectors.

Finally, we comment on the astrophysical relevance of recent
innermost circular orbit (ICO) calculations (for a review,
see~\cite{Baumgarte00a,Blanchet:2001id}).  We note that all current
calculations of ICO configurations for equal mass black holes have
an angular velocity $m \dot{\phi}$ of as low as $0.06$ and as
high as $0.18$.  From Fig.~\ref{fig:circ_phidot}, we see that 
this range corresponds to binary separations of $r = 6 \: m$ or less.
Also, all current ICO calculations are based on the circular
orbit approximation.  However, we have shown in this paper that the
circular orbit approximation has completely broken down for
equal mass black holes with separations of $r \leq 8.5 \: m$.  
This calls into question the astrophysical significance of ICO 
calculations based on circular orbit assumptions.  Certainly, any
gravitational waveforms produced from numerical relativity codes
using initial data that corresponds to 
ICO configurations computed from circular orbit
assumptions will be of little or no astrophysical
significance.

\section{Acknowledgement}
\label{ack}

It is a pleasure to thank Thierry Mora and Clifford Will for
useful discussions and suggestions.
Financial support for this research has been
provided by the 
Jet Propulsion Laboratory (account 100581-A.C.02) under contract with the 
National Aeronautics and Space Administration.
Computational resource support has been provided by the 
NSF NRAC project MCA02N022.


\begin{thebibliography}{29}
\expandafter\ifx\csname natexlab\endcsname\relax\def\natexlab#1{#1}\fi
\expandafter\ifx\csname bibnamefont\endcsname\relax
  \def\bibnamefont#1{#1}\fi
\expandafter\ifx\csname bibfnamefont\endcsname\relax
  \def\bibfnamefont#1{#1}\fi
\expandafter\ifx\csname citenamefont\endcsname\relax
  \def\citenamefont#1{#1}\fi
\expandafter\ifx\csname url\endcsname\relax
  \def\url#1{\texttt{#1}}\fi
\expandafter\ifx\csname urlprefix\endcsname\relax\def\urlprefix{URL }\fi
\providecommand{\bibinfo}[2]{#2}
\providecommand{\eprint}[2][]{\url{#2}}

\bibitem[{\citenamefont{Baumgarte
  et~al.}(1998{\natexlab{a}})\citenamefont{Baumgarte, Cook, Scheel, Shapiro,
  and Teukolsky}}]{Baumgarte98b}
\bibinfo{author}{\bibfnamefont{T.~W.} \bibnamefont{Baumgarte}},
  \bibinfo{author}{\bibfnamefont{G.~B.} \bibnamefont{Cook}},
  \bibinfo{author}{\bibfnamefont{M.~A.} \bibnamefont{Scheel}},
  \bibinfo{author}{\bibfnamefont{S.~L.} \bibnamefont{Shapiro}},
  \bibnamefont{and} \bibinfo{author}{\bibfnamefont{S.~A.}
  \bibnamefont{Teukolsky}}, \bibinfo{journal}{Physical Review D}
  \textbf{\bibinfo{volume}{57}}, \bibinfo{pages}{7299}
  (\bibinfo{year}{1998}{\natexlab{a}}).

\bibitem[{\citenamefont{Baumgarte
  et~al.}(1998{\natexlab{b}})\citenamefont{Baumgarte, Cook, Scheel, Shapiro,
  and Teukolsky}}]{Baumgarte98c}
\bibinfo{author}{\bibfnamefont{T.~W.} \bibnamefont{Baumgarte}},
  \bibinfo{author}{\bibfnamefont{G.~B.} \bibnamefont{Cook}},
  \bibinfo{author}{\bibfnamefont{M.~A.} \bibnamefont{Scheel}},
  \bibinfo{author}{\bibfnamefont{S.~L.} \bibnamefont{Shapiro}},
  \bibnamefont{and} \bibinfo{author}{\bibfnamefont{S.~A.}
  \bibnamefont{Teukolsky}}, \bibinfo{journal}{Physical Review D}
  \textbf{\bibinfo{volume}{57}}, \bibinfo{pages}{6181}
  (\bibinfo{year}{1998}{\natexlab{b}}).

\bibitem[{\citenamefont{Bonazzola et~al.}(1997)\citenamefont{Bonazzola,
  Gourgoulhon, and Marck}}]{Bonazzola97}
\bibinfo{author}{\bibfnamefont{S.}~\bibnamefont{Bonazzola}},
  \bibinfo{author}{\bibfnamefont{E.}~\bibnamefont{Gourgoulhon}},
  \bibnamefont{and} \bibinfo{author}{\bibfnamefont{J.-A.} \bibnamefont{Marck}},
  \bibinfo{journal}{Phys. Rev. D} \textbf{\bibinfo{volume}{56}},
  \bibinfo{pages}{7740} (\bibinfo{year}{1997}).

\bibitem[{\citenamefont{Gourgoulhon
  et~al.}(2001{\natexlab{a}})\citenamefont{Gourgoulhon, Grandclement,
  Taniguchi, Marck, and Bonazzola}}]{Gourgoulhon01}
\bibinfo{author}{\bibfnamefont{E.}~\bibnamefont{Gourgoulhon}},
  \bibinfo{author}{\bibfnamefont{P.}~\bibnamefont{Grandclement}},
  \bibinfo{author}{\bibfnamefont{K.}~\bibnamefont{Taniguchi}},
  \bibinfo{author}{\bibfnamefont{J.}~\bibnamefont{Marck}}, \bibnamefont{and}
  \bibinfo{author}{\bibfnamefont{S.}~\bibnamefont{Bonazzola}},
  \bibinfo{journal}{Phys. Rev. D} \textbf{\bibinfo{volume}{63}},
  \bibinfo{pages}{064029} (\bibinfo{year}{2001}{\natexlab{a}}).

\bibitem[{\citenamefont{Marronetti et~al.}(1999)\citenamefont{Marronetti,
  Mathews, and Wilson}}]{Marronetti:1999ya}
\bibinfo{author}{\bibfnamefont{P.}~\bibnamefont{Marronetti}},
  \bibinfo{author}{\bibfnamefont{G.~J.} \bibnamefont{Mathews}},
  \bibnamefont{and} \bibinfo{author}{\bibfnamefont{J.~R.}
  \bibnamefont{Wilson}}, \bibinfo{journal}{Phys. Rev.}
  \textbf{\bibinfo{volume}{D60}}, \bibinfo{pages}{087301}
  (\bibinfo{year}{1999}).

\bibitem[{\citenamefont{Shibata}(1998)}]{Shibata98}
\bibinfo{author}{\bibfnamefont{M.}~\bibnamefont{Shibata}},
  \bibinfo{journal}{Phys. Rev. D} \textbf{\bibinfo{volume}{58}},
  \bibinfo{pages}{024012} (\bibinfo{year}{1998}).

\bibitem[{\citenamefont{Teukolsky}(1998)}]{Teukolsky98}
\bibinfo{author}{\bibfnamefont{S.}~\bibnamefont{Teukolsky}},
  \bibinfo{journal}{ApJ} \textbf{\bibinfo{volume}{504}}, \bibinfo{pages}{442}
  (\bibinfo{year}{1998}).

\bibitem[{\citenamefont{Yo et~al.}(2001)\citenamefont{Yo, Baumgarte, and
  Shapiro}}]{Yo01}
\bibinfo{author}{\bibfnamefont{H.-J.} \bibnamefont{Yo}},
  \bibinfo{author}{\bibfnamefont{T.}~\bibnamefont{Baumgarte}},
  \bibnamefont{and} \bibinfo{author}{\bibfnamefont{S.}~\bibnamefont{Shapiro}},
  \bibinfo{journal}{Phys. Rev. D} \textbf{\bibinfo{volume}{63}}
  (\bibinfo{year}{2001}).

\bibitem[{\citenamefont{Duez et~al.}(2001)\citenamefont{Duez, Baumgarte, and
  Shapiro}}]{Duez00}
\bibinfo{author}{\bibfnamefont{M.~D.} \bibnamefont{Duez}},
  \bibinfo{author}{\bibfnamefont{T.~W.} \bibnamefont{Baumgarte}},
  \bibnamefont{and} \bibinfo{author}{\bibfnamefont{S.~L.}
  \bibnamefont{Shapiro}}, \bibinfo{journal}{Phys. Rev.}
  \textbf{\bibinfo{volume}{D63}}, \bibinfo{pages}{084030}
  (\bibinfo{year}{2001}).

\bibitem[{\citenamefont{Duez et~al.}(2002)\citenamefont{Duez, Baumgarte,
  Shapiro, Shibata, and Uryu}}]{Duez02}
\bibinfo{author}{\bibfnamefont{M.~D.} \bibnamefont{Duez}},
  \bibinfo{author}{\bibfnamefont{T.~W.} \bibnamefont{Baumgarte}},
  \bibinfo{author}{\bibfnamefont{S.~L.} \bibnamefont{Shapiro}},
  \bibinfo{author}{\bibfnamefont{M.}~\bibnamefont{Shibata}}, \bibnamefont{and}
  \bibinfo{author}{\bibfnamefont{K.}~\bibnamefont{Uryu}},
  \bibinfo{journal}{Phys. Rev.} \textbf{\bibinfo{volume}{D65}},
  \bibinfo{pages}{024016} (\bibinfo{year}{2002}).

\bibitem[{\citenamefont{Gourgoulhon
  et~al.}(2001{\natexlab{b}})\citenamefont{Gourgoulhon, Grandclement, and
  Bonazzola}}]{Gourgoulhon:2001ec}
\bibinfo{author}{\bibfnamefont{E.}~\bibnamefont{Gourgoulhon}},
  \bibinfo{author}{\bibfnamefont{P.}~\bibnamefont{Grandclement}},
  \bibnamefont{and}
  \bibinfo{author}{\bibfnamefont{S.}~\bibnamefont{Bonazzola}},
  \bibinfo{journal}{Phys. Rev. D} \textbf{\bibinfo{volume}{65}},
  \bibinfo{pages}{044020} (\bibinfo{year}{2001}{\natexlab{b}}).

\bibitem[{\citenamefont{Grandclement et~al.}(2001)\citenamefont{Grandclement,
  Gourgoulhon, and Bonazzola}}]{Grandclement:2001ed}
\bibinfo{author}{\bibfnamefont{P.}~\bibnamefont{Grandclement}},
  \bibinfo{author}{\bibfnamefont{E.}~\bibnamefont{Gourgoulhon}},
  \bibnamefont{and}
  \bibinfo{author}{\bibfnamefont{S.}~\bibnamefont{Bonazzola}},
  \bibinfo{journal}{Phys. Rev. D} \textbf{\bibinfo{volume}{65}},
  \bibinfo{pages}{044021} (\bibinfo{year}{2001}).

\bibitem[{\citenamefont{Friedman et~al.}(2002)\citenamefont{Friedman, Uryu, and
  Shibata}}]{Friedman02}
\bibinfo{author}{\bibfnamefont{J.~L.} \bibnamefont{Friedman}},
  \bibinfo{author}{\bibfnamefont{K.}~\bibnamefont{Uryu}}, \bibnamefont{and}
  \bibinfo{author}{\bibfnamefont{M.}~\bibnamefont{Shibata}},
  \bibinfo{journal}{Phys. Rev. D} \textbf{\bibinfo{volume}{65}},
  \bibinfo{pages}{064035} (\bibinfo{year}{2002}).

\bibitem[{\citenamefont{Cook}(1994)}]{Cook94}
\bibinfo{author}{\bibfnamefont{G.~B.} \bibnamefont{Cook}},
  \bibinfo{journal}{Phys. Rev. D} \textbf{\bibinfo{volume}{50}},
  \bibinfo{pages}{5025} (\bibinfo{year}{1994}).

\bibitem[{\citenamefont{Baumgarte}(2000)}]{Baumgarte00a}
\bibinfo{author}{\bibfnamefont{T.~W.} \bibnamefont{Baumgarte}},
  \bibinfo{journal}{Phys. Rev. D} \textbf{\bibinfo{volume}{62}},
  \bibinfo{pages}{024018} (\bibinfo{year}{2000}).

\bibitem[{\citenamefont{Blanchet et~al.}(1998)\citenamefont{Blanchet, Faye, and
  Ponsot}}]{Blanchet98}
\bibinfo{author}{\bibfnamefont{L.}~\bibnamefont{Blanchet}},
  \bibinfo{author}{\bibfnamefont{G.}~\bibnamefont{Faye}}, \bibnamefont{and}
  \bibinfo{author}{\bibfnamefont{B.}~\bibnamefont{Ponsot}},
  \bibinfo{journal}{Phys. Rev.} \textbf{\bibinfo{volume}{D58}},
  \bibinfo{pages}{124002} (\bibinfo{year}{1998}).

\bibitem[{\citenamefont{Itoh et~al.}(2001)\citenamefont{Itoh, Futamase, and
  Asada}}]{Itoh01}
\bibinfo{author}{\bibfnamefont{Y.}~\bibnamefont{Itoh}},
  \bibinfo{author}{\bibfnamefont{T.}~\bibnamefont{Futamase}}, \bibnamefont{and}
  \bibinfo{author}{\bibfnamefont{H.}~\bibnamefont{Asada}},
  \bibinfo{journal}{Phys. Rev. D} \textbf{\bibinfo{volume}{63}},
  \bibinfo{pages}{064038} (\bibinfo{year}{2001}).

\bibitem[{\citenamefont{Pati and Will}(2002)}]{Pati02}
\bibinfo{author}{\bibfnamefont{M.~E.} \bibnamefont{Pati}} \bibnamefont{and}
  \bibinfo{author}{\bibfnamefont{C.~M.} \bibnamefont{Will}},
  \bibinfo{journal}{Phys. Rev. D} \textbf{\bibinfo{volume}{65}},
  \bibinfo{pages}{104008} (\bibinfo{year}{2002}).

\bibitem[{\citenamefont{Damour et~al.}(2000)\citenamefont{Damour, Jaranowski,
  and Sch\"afer}}]{Damour2000_Poinc}
\bibinfo{author}{\bibfnamefont{T.}~\bibnamefont{Damour}},
  \bibinfo{author}{\bibfnamefont{P.}~\bibnamefont{Jaranowski}},
  \bibnamefont{and}
  \bibinfo{author}{\bibfnamefont{G.}~\bibnamefont{Sch\"afer}},
  \bibinfo{journal}{Phys. Rev. D} \textbf{\bibinfo{volume}{62}},
  \bibinfo{pages}{021501} (\bibinfo{year}{2000}).

\bibitem[{\citenamefont{Blanchet and Faye}(2001)}]{Blanchet01a}
\bibinfo{author}{\bibfnamefont{L.}~\bibnamefont{Blanchet}} \bibnamefont{and}
  \bibinfo{author}{\bibfnamefont{G.}~\bibnamefont{Faye}},
  \bibinfo{journal}{Phys. Rev.} \textbf{\bibinfo{volume}{D63}},
  \bibinfo{pages}{062005} (\bibinfo{year}{2001}).

\bibitem[{\citenamefont{Gopakumar et~al.}(1997)\citenamefont{Gopakumar, Iyer,
  and Iyer}}]{Gopakumar97}
\bibinfo{author}{\bibfnamefont{A.}~\bibnamefont{Gopakumar}},
  \bibinfo{author}{\bibfnamefont{B.}~\bibnamefont{Iyer}}, \bibnamefont{and}
  \bibinfo{author}{\bibfnamefont{S.}~\bibnamefont{Iyer}},
  \bibinfo{journal}{Phys. Rev. D} \textbf{\bibinfo{volume}{55}},
  \bibinfo{pages}{6030} (\bibinfo{year}{1997}).

\bibitem[{\citenamefont{Lincoln and Will}(1990)}]{Lincoln90}
\bibinfo{author}{\bibfnamefont{C.~W.} \bibnamefont{Lincoln}} \bibnamefont{and}
  \bibinfo{author}{\bibfnamefont{C.~M.} \bibnamefont{Will}},
  \bibinfo{journal}{Phys. Rev. D} \textbf{\bibinfo{volume}{42}},
  \bibinfo{pages}{1123} (\bibinfo{year}{1990}).

\bibitem[{\citenamefont{Mora and Will}(2002)}]{Mora02}
\bibinfo{author}{\bibfnamefont{T.}~\bibnamefont{Mora}} \bibnamefont{and}
  \bibinfo{author}{\bibfnamefont{C.~M.} \bibnamefont{Will}},
  \bibinfo{journal}{Phys. Rev. D} \textbf{\bibinfo{volume}{66}},
  \bibinfo{pages}{101501(R)} (\bibinfo{year}{2002}).

\bibitem[{\citenamefont{Epstein and Wagoner}(1975)}]{Epstein75}
\bibinfo{author}{\bibfnamefont{R.}~\bibnamefont{Epstein}} \bibnamefont{and}
  \bibinfo{author}{\bibfnamefont{R.~V.} \bibnamefont{Wagoner}},
  \bibinfo{journal}{ApJ} \textbf{\bibinfo{volume}{197}},
  \bibinfo{pages}{717} (\bibinfo{year}{1975}).

\bibitem[{\citenamefont{Wagoner and Will}(1976)}]{Wagoner76}
\bibinfo{author}{\bibfnamefont{R.~V.} \bibnamefont{Wagoner}} \bibnamefont{and}
  \bibinfo{author}{\bibfnamefont{C.~M.} \bibnamefont{Will}},
  \bibinfo{journal}{ApJ} \textbf{\bibinfo{volume}{210}},
  \bibinfo{pages}{764} (\bibinfo{year}{1976}).

\bibitem[{\citenamefont{Turner and Will}(1978)}]{Turner78}
\bibinfo{author}{\bibfnamefont{M.}~\bibnamefont{Turner}} \bibnamefont{and}
  \bibinfo{author}{\bibfnamefont{C.~M.} \bibnamefont{Will}},
  \bibinfo{journal}{ApJ} \textbf{\bibinfo{volume}{220}},
  \bibinfo{pages}{1107} (\bibinfo{year}{1978}).

\bibitem[{\citenamefont{Miller and Suen}(2003)}]{Miller03b}
\bibinfo{author}{\bibfnamefont{M.}~\bibnamefont{Miller}} \bibnamefont{and}
  \bibinfo{author}{\bibfnamefont{W.-M.} \bibnamefont{Suen}}
  (\bibinfo{year}{2003}), \bibinfo{note}{submitted}.

\bibitem[{\citenamefont{Cook and Scheel}(1997)}]{Cook97b}
\bibinfo{author}{\bibfnamefont{G.~B.} \bibnamefont{Cook}} \bibnamefont{and}
  \bibinfo{author}{\bibfnamefont{M.~A.} \bibnamefont{Scheel}},
  \bibinfo{journal}{Phys. Rev. D} \textbf{\bibinfo{volume}{56}},
  \bibinfo{pages}{4775} (\bibinfo{year}{1997}).

\bibitem[{\citenamefont{Blanchet}(2002)}]{Blanchet:2001id}
\bibinfo{author}{\bibfnamefont{L.}~\bibnamefont{Blanchet}},
  \bibinfo{journal}{Phys. Rev.} \textbf{\bibinfo{volume}{D65}},
  \bibinfo{pages}{124009} (\bibinfo{year}{2002}).

\end{thebibliography}


\end{document}